# The Aligned Nuclear Targets for Investigation of Time Reversal Invariance Violation: Thermal Heating and Optimization of Target Dimensions.

A.G.Beda, A.S. Gerasimov
Institute for Theoretical and Experimental Physics, Moscow.

The thermal heating of aligned nuclear targets of HIO$_3$, LiIO$_3$ and Sb target materials under neutron irradiation at JSNS is considered. It is shown that presently the targets of large volumes – several tens of cm$^3$ - can be used in experiment. The optimal target dimensions are recommended for investigation with resonance neutrons. The use of proposed aligned targets at the new neutron spallation source JSNS ( Japan) will make possible to discover TRIV or decrease the present limit on the intensity of parity conserving time violating interaction by two-three order of magnitude.

## 1. Introduction

The investigation of TRIV in the neutron- nuclei interactions is very perspective due to large enhancement of the effects of violation in the compound resonances of nuclei in comparison with its values in elementary nucleon-nucleon interactions. One needs the polarized neutron beam and oriented (polarized or aligned ) nuclear targets for these investigations. Although the first proposals on the investigation of TRIV in the neutron-nuclei interactions have been made more than twenty years ago [1-4] real progress in this field was absent for the lack of appropriate oriented nuclear targets. Recently the appropriate target materials of HIO$_3$, LiIO$_3$ and Sb single crystals were proposed in [5]. In these materials the I and Sb nuclei can be aligned by brute force method at millikelvin temperatures. The target geometry which provides possibility of necessary cooling of target material was proposed in [6]. One must use targets of large volume (several tens of cm$^3$) to achieve high statistical accuracy of measurements. However, it should be noted that the cooling of large volume targets irradiated by intense neutron beam down to millikelvin temperatures is a challenge problem because of large thermal heating of target material caused by neutron-nuclei interactions. The problem of thermal heating and optimization of target geometry are considered below.

## 2. The thermal heating of the targets

We shall consider the problem of the thermal heating of large volume targets taking into account the recommendations of [6]. According to these recommendations the single crystals of HIO$_3$, LiIO$_3$ and Sb must be cut into plates with thickness not more 1 cm to provide effective cooling by $^3$He-$^4$He mixture circulating between the plates. The calculations were performed for the targets of HIO$_3$, LiIO$_3$ and Sb with volume V = 5·5·5 cm$^3$ (five plates 5·5·1 cm$^3$ with gaps 0.5 cm between the plates, see fig. 1). The mixture of $^3$He-$^4$He is 6,4% $^3$He and 93,6% $^4$He. It was assumed that the target are placed at the distance of 10 m from the center of resonance neutron source JSNS. The polarized neutron flux density at the target is

$$\Phi(E)\,dE = N_0\,k\,f\,\Omega\frac{dE}{E} = \Phi_0\,\frac{dE}{E} \quad , \tag{1}$$

where the constant $N_0 = 2 \cdot 10^{12}$ for the neutron spallation source LANCE (at proton energy $E_p$=800 MeV and proton current $I_p$= 70 µA) [7], $N_0$ will be 20 times more for JSNS since in last case $E_p$=3000 MeV and $I_p$= 333 µA, k =0.14 is a polarizer transmission , f ~0.63 is the fraction of moderator surface viewed by the target through the collimator. So hereafter we will use formula (1) for calculations where $\Phi_0 = 3.6 \cdot 10^6$ for JSNS. The thermal heating will be determined as product of number of neutron- nuclei (nucleus of He, H, Li, I and Sb) interactions in the target material by the energy generated in these interactions.

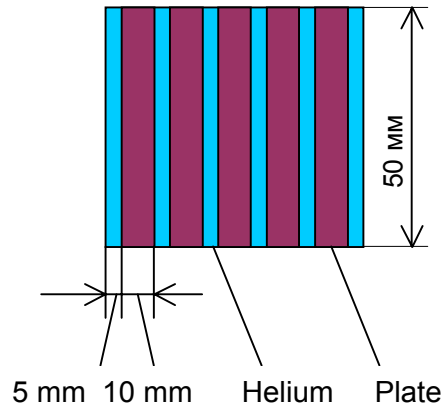

Fig.1 Schematic of the target.

## 3. The thermal heating of He

As noted above the liquid helium cooling the plates is a mixture of $^3$He-$^4$He. At that $^3$He is a high absorbing nuclide. The cross-section of the $^3$He(n,p)$^3$H reaction for thermal neutrons ($E_0$ = 0.0235 eV) is $\sigma_0$ = 5333 bn [8] and depends on energy as $1/V$. The energy generated in this reaction is 0.764 MeV ($1.2 \cdot 10^{-13}$ J) and it absorbed by target near the point of interaction.

The number of the interactions of neutrons with $^3$He is

$$<N_3 \sigma \Phi> = N_3 \int_{E_1}^{E_2} \sigma(E) \Phi_0 \frac{dE}{E} = N_3 \int_{E_1}^{E_2} \sigma_0 \sqrt{\frac{E_0}{E}} \Phi_0 \frac{dE}{E} = 0.318 N_3 \sigma_0 \Phi_0 \quad (2)$$

where $N_3 = 8.81 \cdot 10^{22}$ is total number of $^3$He nuclei, $E_0$ = 0,0253 eV, $E_1$ = 1 eV, $E_2$ = 1000 eV.

The total number of interactions is

$$<N_3 \sigma \Phi> = 0.318 \cdot 8.81 \cdot 10^{22} \cdot 5333 \cdot 10^{-24} \cdot 3.6 \cdot 10^6 = 5.4 \cdot 10^8 \quad (3)$$

The number of the neutrons incident upon helium gaps between the plates is

$$<\Phi S> = \Phi_0 S \int_1^{1000} \frac{dE}{E} = (3.6 \cdot 10^6 \, n/cm^2 s) \times (15 \, cm^2) \times \ln 1000 = 3.7 \cdot 10^8 \, n/s \quad (4)$$

The comparison of the number of the neutron interactions with $^3$He - $<N_3 \sigma \Phi>$ with number of neutrons incident upon the helium between the plates - $<\Phi S>$ demonstrates that the number of interactions is greater that the number of incident neutrons ! It means that one must take into account the depression of neutron flux at transmission through the target. If one take into account this depression of neutron flux along the length of target then the number of the interactions of neutrons with $^3$He will be $<< N_3 \sigma \Phi >> = 2.2 \cdot 10^8$ 1/s.

The thermal heating is calculated as product of number of (n,$^3$He) interactions by the energy generated in these interactions

$$<Q_3> = Q_3 \cdot <<\sigma \Phi S>> = (1.2 \cdot 10^{-13} \text{ J}) \cdot 2.2 \cdot 10^8 = 270 \cdot 10^{-7} \text{ W} = 27 \text{ }\mu\text{W} \qquad (5)$$

Hence the thermal heating in the target due to neutron capture by $^3$He is 27 µW.

The $^4$He nuclide does not absorb the neutrons and its contribution in the thermal heating can be caused only by slowing-down of the neutrons. But in the case of $^4$He this contribution can be neglected.

## 4. The thermal heating of HIO$_3$

### 1) H

The $^1$H hydrogen is a low absorbing nuclide. The cross-section of the $^1$H (n,γ)D reaction for thermal neutrons ($E_0 = 0.0235$ eV) is $\sigma_0 = 0.33$ bn [8]. The energy of γ quanta being emitted at the neutron capture by $^1$H is 2.23 MeV ($3.5 \cdot 10^{-13}$ J). The value of the thermal heating in the target due to $^1$H (n,γ)D reaction resulting from formulas (1-2) is $Q_{1H} = 7$ µW (for $N_1 = 1.96 \cdot 10^{24}$). If one take into account the probability of γ quanta escapement from the target this value will fall to 4 µW. As for as thermal heating concerned, the neutron slowing down by hydrogen is very important since the mean energy transmitted to proton in each collision with neutron is equal to the half of the incident neutron energy. The rate of elastic scattering reactions in the energy range dE is

$$dN = \sigma_s \Phi_0 N_1 \frac{dE}{E} \qquad (6)$$

where $\sigma_s$ is a cross-section of neutron elastic scattering by $^1$H hydrogen. The thermal heating in the energy range dE is

$$dQ_{nH} = \frac{E}{2} dN = \sigma_s \Phi_0 N_1 \frac{E}{2} \frac{dE}{E} = \sigma_s \Phi_0 N_1 \frac{dE}{2} \qquad (7)$$

The total thermal heating over all energy spectrum of neutrons is

$$Q_{nH} = \sigma_s \Phi_0 N_1 \frac{E}{2} \qquad (8)$$

Note that the thermal heating value is proportional to the upper limit of energy spectrum of incident neutrons. The cross-section $\sigma_s$ is constant at neutron energy below 20 keV. It decreases at higher energy. Taking into account this dependence the thermal heating of target due to the neutron elastic scattering by hydrogen can be estimated as 35 µW at upper limit E = 5 MeV.

### 2) I

Thermal heating of target in the case of I target is determined by $^{127}$I (n,γ) $^{128}$I reaction. In the process the part of energy is emitted in the form of prompt gamma radiation and other – in the form of radioactive decay energy of $^{128}$I (γ- and β- radiations). A cross-section of (n,γ) reaction in the resonance range of energy includes one part depending on energy as $1/V$ and other part determined by resonances. The number of reactions determined by first part is

$$<N_{127}\sigma_{(1/V)}\Phi> = 0.318 \cdot \sigma_0 \cdot \Phi_0 \cdot N_{127} \qquad (9)$$

where $\sigma_0 = 6.2$ bn, $N_{127} = 1.96 \cdot 10^{24}$.

The number of reactions determined by resonances is

$$<N_7 \sigma_{rez} \Phi> = \Phi_0 N_7 \sum_k \Delta I_k \qquad (10)$$

The $^{127}$I nuclide has a large number of neutron resonances but in the low energy region there are about ten strong resonances [8]. Their contribution in total cross-section is $\Sigma \Delta I_K = 9.39$ bn. Hence the total number of neutron interactions with $^{127}$I nuclides is

$$<N_{127}\sigma_{127}\Phi> = \Phi_0 \cdot N_{127}(0.318 \cdot \sigma_0 + \Sigma \Delta I_K) = 3.6 \cdot 10^6 \cdot 1.96 \cdot 10^{24} \cdot 11.4 \cdot 10^{-24} = 8 \cdot 10^7 \qquad (11)$$

The thermal heating due to the radioactive decay of $^{128}$I is (see Table 1)

$$Q_\beta = 0.831 \cdot 1{,}6 \cdot 10^{-13} \cdot 8 \cdot 10^7 = 11 \ \mu W \qquad (12)$$

Table 1. The radioactive decay energy of $^{127}$I, $^{121}$Sb and $^{123}$Sb nuclides [9].

| Nuclide | Type of radiation | Energy (MeV) |
|---|---|---|
| $^{128}$I | Gamma and X-ray | 0.085 |
|  | Beta | 0.747 |
|  | Total | 0.831 |
| $^{122}$Sb | Gamma and X-ray | 0.44 |
|  | Beta | 0.56 |
|  | Total | 1.00 |
| $^{124}$Sb | Gamma and X-ray | 1.8 |
|  | Beta | 0.38 |
|  | Total | 2.18 |

One must take into account the energy spectrum of prompt γ-gamma radiation at the calculation of the thermal heating due to the prompt γ-radiation. The approximate energy spectra of prompt γ-radiation of I and Sb nuclei (for natural abundance) are shown in Table 2. For calculation of thermal heating one must determine the mean energy of prompt γ-quanta to one neutron capture and to consider the probability of its escape from the target, then the mean energy will be

$$\overline{E} = \frac{1}{100} \sum_k E_k n_k \left(1 - e^{-\mu l}\right)_k \qquad (13)$$

where $E_k$ is mean energy of prompt γ-quanta in the energy range, $n_k$ is a number of γ-quanta and μ is a absorption coefficient. Then the mean energy of prompt γ quanta in the case of $^{127}$I(n,γ) reaction is

$$\overline{E} = 1.317 \text{ MeV}, \qquad (14)$$

Table 2. Number of γ-quanta in different energy ranges to (n,γ) 100 reactions [10].

| Nuclei | 0 – 1 MeV | 1 – 2 MeV | 2 – 3 MeV | 3 – 5 MeV | 5 – 7 MeV | 7 – 9 MeV |
|---|---|---|---|---|---|---|
| Sb | 150 | 99 | 58 | 36 | 12 | - |
| J | >30 | - | - | 61 | 25 | - |

For calculation of thermal heating one must determine the mean energy of prompt γ-quanta to one neutron capture and to consider the probability of its escape from the target, then the mean energy will be

$$\overline{E} = \frac{1}{100} \sum_k E_k n_k \left(1 - e^{-\mu l}\right)_k \qquad (15)$$

where $E_k$ is mean energy of prompt γ-quanta in the energy range, $n_k$ is a number of γ-quanta and μ is an absorption coefficient. Then the mean energy of prompt γ-quanta in the case of $^{127}$I(n,γ) reaction is

$$\overline{E} = 1.317 \text{ MeV}, \qquad (16)$$

The total thermal heating due to prompt γ-radiation originating from $^{127}$I (n,γ) $^{128}$I reaction is

$$Q_\gamma = <N_{127}\sigma_{127}\Phi> \overline{E_\gamma} = 8 \cdot 10^7 \cdot 1.317 \cdot 1.6 \cdot 10^{-13} = 17 \text{ μW} \qquad (17)$$

3) O

The number of the oxygen nuclei in the target of HIO$_3$ is 3 times more than number of the $^{127}$I nuclei. However, the cross-section of neutron absorption by $^{16}$O nuclei is very small ( 0.19 mbn ) and its contribution in thermal heating can be neglected.

## 5. The thermal heating of LiIO$_3$

1) Li

The lithium incorporating in target plates involves two nuclides: $^6$Li (7.5%) and $^7$Li (92.5%). The cross-section of (n,γ) reaction for both nuclides is very small ( 0.03 bn [8] ) so its contribution in thermal heating of the target can be neglected. At the same time the cross-section of $^6$Li (n,α)$^3$H reaction is very large and its value for thermal neutrons (at $E_0 = 0.0235$ eV) is $\sigma_0 = 940$ bn [8] and depends on energy as $1/V$ (where $V$ is velocity of the neutron). The reaction energy is E= 4.785 MeV = $7.65 \cdot 10^{-13}$ J. Since the path of the α-particle is rather short the whole energy is absorbed near the point the neutron absorption and calculation of the thermal heating due to $^6$Li (n,α)$^3$H reaction is a relatively simple. However, one must take into account the depression of the neutron flux resulted from large cross-section of this reaction. Let us compare the number of incident neutrons with number of neutrons absorbed by $^6$Li. The number of the neutrons incident upon the plates is

$$<\Phi S> = \Phi_0 S \int_1^{1000} \frac{dE}{E} = (3.6 \cdot 10^6 \, n/cm^2 s) \times (25 \, cm^2) \times \ln 1000 = 6.2 \cdot 10^8 \, n/s \quad (18)$$

The number of absorbed neutrons is

$$<N_6 \sigma \Phi> = 1,5 \cdot 10^8 \, n/s \quad (19)$$

Hence the 24% of incident neutrons is absorbed in target. In the first approximation one can consider that the neutron flux density decreases linearly on the target length and use in the calculation mean value of neutron flux equal 88,8% of incident neutron flux. Then the number of absorbed neutrons with consideration for neutron flux depression is $<N_6 \sigma \Phi> = 1.3 \cdot 10^8$ and the value of thermal heating is

$$Q_6 = <N_6 \sigma_6 \Phi> E_\alpha \, 0.875 = 1.39 \cdot 10^{23} \cdot 940 \cdot 10^{-24} \cdot 3.6 \cdot 10^6 \cdot 0.318 \cdot 7.65 \cdot 10^{-13} \cdot 0.875 = 100 \, \mu W \quad (20)$$

2) I

The calculation of the thermal heating due to $^{127}$I (n,γ) $^{128}$I reaction for the target of LiIO$_3$ single crystal plates is analogous to the case of HIO$_3$ single crystal plates . However one must correct for the difference of $^{127}$I nuclei in the targets ($1.96 \cdot 10^{24}$ for HIO$_3$ target and $1,86 \cdot 10^{24}$ for LiIO$_3$ target) and for depression of neutron flux in the case of LiIO$_3$ target. Then the thermal heating due to the prompt gamma radiation energy will be 14 μW and due to the radioactive decay energy - 11 μW.

## 6. The thermal heating of Sb

Now we shall consider the problem of thermal heating of target with plates of Sb with natural abundance ( 57.3% $^{121}$Sb, $\sigma_0$ =5.9 bn and 42.7% $^{123}$Sb, $\sigma_0$ =4.15 bn [8]). The thermal heating of target due to Sb(n,γ) reaction also caused by prompt gamma radiation and radioactive decay of reaction products. At the beginning we shall calculate the number of (n,γ) reactions in each nuclide of Sb. Since each of them has strong resonances one must take account of their contribution in total cross-section of reaction with use of formula (11). Then the total cross-section for $^{121}$Sb is

$$\left(0.318\sigma_0 + \sum_k \Delta I_k\right) = 0.318 \cdot 5.9 \, 10^{-24} + 11.3 \cdot 10^{-24} = 13.2 \cdot 10^{-24} \quad (21)$$

for $^{123}$Sb

$$\left(0.318\sigma_0 + \sum_k \Delta I_k\right) = 0.318 \cdot 4.15 \cdot 10^{-24} + 5.7 \cdot 10^{-24} = 7 \cdot 10^{-24} \quad (22)$$

The number of (n,γ) reactions for each Sb nuclide is

for $^{121}$Sb
$$< N_{121}\sigma_{121}\Phi > = 2.37 \cdot 10^{24} \cdot 13.2 \cdot 10^{-24} \cdot 3.6 \cdot 10^{6} = 1.13 \cdot 10^{8} \qquad (23)$$

for $^{123}$Sb
$$< N_{123}\sigma_{123}\Phi > = 1.77 \cdot 10^{24} \cdot 7 \cdot 10^{-24} \cdot 3.6 \cdot 10^{6} = 4.45 \cdot 10^{7} \qquad (24)$$

The mean energy of prompt γ-radiation on one reaction calculated with the formula (15) is

$$\overline{E_{\gamma}} = 2.7 \text{ MeV} \qquad (25)$$

The total thermal heating due of prompt gamma radiation originating from Sb (n,γ) reaction is

$$Q_{\gamma} = [< N_{121}\sigma_{121}\Phi > + < N_{123}\sigma_{123}\Phi >] \overline{E_{\gamma}} = 1.57 \cdot 10^{8} \cdot 2.7 \cdot 1.6 \cdot 10^{-13} = 68 \text{ μW} \qquad (26)$$

In calculation of thermal heating due to the radioactive decay we assume that that $^{122}$Sb and $^{124}$Sb radioactive nuclei decay immediately after production though their half-lives are : 2.7 days for $^{122}$Sb and 60 days for $^{124}$Sb. Then one can calculate thermal heating due to the radioactive decay using the data of Table 2. The thermal heating in the case of $^{122}$Sb (with consideration for escape in part of gamma radiation from target ) is

$$<Q_{Sb1}> = 1.13 \cdot 10^{8} \cdot 1.60 \cdot 10^{-13} = 18 \text{ μW} \qquad (27)$$

Analogous quantity for $^{124}$Sb is

$$<Q_{Sb3}> = 4.45 \cdot 10^{7} \cdot 3.49 \cdot 10^{-13} = 15.5 \text{ μW} \qquad (28)$$

## 7. The thermal heating due to the neutron elastic scattering

At the neutron-nuclei elastic scattering neutron transmits part of its energy to the nuclei then this energy changes into the heat. This process is slowing –down for the neutrons and heating for the target. The thermal heating due to the neutron elastic scattering can be estimated by the following way. The mean energy loss $\Delta E$ of neutron at elastic scattering is

$$\frac{\Delta E}{E} = \frac{2A}{(A+1)^{2}}(1 - \overline{\cos \varphi}), \qquad (29)$$

where E is a neutron energy, A is nucleus mass, $\overline{\cos \varphi}$ is mean value of cosine of scattering angle in the centre mass system (C.M.S.) of neutron and nuclei. The neutron scattering is isotropic in C.M.S. at neutron energy smaller 100 keV, under these conditions $\overline{\cos \varphi} = 0$. Then

$$\frac{\Delta E}{E} = \frac{2A}{(A+1)^{2}}. \qquad (30)$$

The thermal heating in the target due to the neutron slowing-down is

$$Q = N \int_{E_1}^{E_2} \Delta E \sigma_s(E) \Phi(E) dE \qquad (31)$$

where $N$ is a number of scattering nuclei, $\sigma_s$ is an elastic scattering cross-section. The Q value can be obtained by substitution $\Delta E$ of (28) and $\Phi(E)dE$ of (1). The result is

$$Q = \frac{2A}{(A+1)^2} N \sigma_s \Phi_0 \int_{E_1}^{E_2} dE = \frac{2A}{(A+1)^2} N \sigma_s \Phi_0 (E_2 - E_1) = \frac{2A}{(A+1)^2} N \sigma_s \Phi_0 E_2. \qquad (32)$$

Here it was assumed that an elastic scattering cross-section does not depend on energy and that $E_2 \gg E_1$. The numerical results of above calculations obtained at $E_2 = 100$ кeV and $\Phi_0 E_2 = 8.064 \cdot 10^{-10}$ J/cm²·s are presented in the Table 3.

The total thermal heating in the target of LiIO$_3$ due to the neutron elastic scattering is of 0.162 µW ( about 0,1% from total thermal heating), in the target of Sb – 0.02 µW (about 0.02% from total thermal heating). One can see that the thermal heating due to the neutron elastic scattering in the above targets can be neglected. However, in the case of HIO$_3$ this contribution in the total thermal heating is dominating – 35 µW for $E_2 = 5$ MeV (at total thermal heating 62.6 µW ) since the average energy transmitted to proton in each collision with neutron is equal to the half of the incident neutron energy.

Table 3. The thermal heating due to the neutron elastic scattering

| Nucleus | A | $2A/(A+1)^2$ | $\sigma_s$, $10^{-24}$ см² | $n$, $10^{24}$ nuclei | Q, µW |
|---|---|---|---|---|---|
| $^3$He | 3 | 0.375 | 3.10 | 0.0881 | 0.006 |
| $^4$He | 4 | 0.32 | 0.76 | 1.288 | 0.02 |
| $^6$Li | 6 | 0.245 | - | 0.139· | - |
| $^7$Li | 7 | 0.219 | 0.97 | 1.72 | 0.02 |
| $^{16}$O | 16 | 0.111 | 3.76 | 5.58 | 0.135 |
| $^{27}$Al | 27 | 0.0689 | 1.41 | 1.44 | 0.008 |
| Sb | 121.75 | 0.0162 | 6.0 | 4.13 | 0.02 |
| $^{127}$I | 127 | 0.0155 | 4.2 | 1.86 | 0.007 |

The thermal heating of targets due to the all processes considered above is sown in Table 4. Recall that the thermal heating in targets shown in Table 4 was calculated for targets of V = 5·5·5 cm³ volume. Presently the targets of such proportions can be produced from Sb, HIO$_3$ and LiIO$_3$ single crystals [5]. The Sb$^{121}$, Sb$^{123}$ and I$^{127}$ nuclei can be aligned in these single crystals by brute force method using the modern dilution refrigerator that is precooled by two stage pulse-tube refrigerator without any cryoliquids [11].

Table 4. Thermal heating of targets

| Compound, Nuclide | Number of nuclei in target, $10^{24}$ nuclei | Reaction | Energy of reaction, MeV | Number or reaction in target, 1/s | Thermal heating μW |
|---|---|---|---|---|---|
| **Helium** | 1.376 | | | | |
| $^3$He | $8.81 \cdot 10^{-2}$ | (n,p) | 0.764 | $2.2 \cdot 10^8$ | **27** |
| $^4$He | 1.288 | | | | |
| **LiIO$_3$** | 1.86 | | | | |
| $^6$Li | 0.14 | (n,α) | 4.785 | $1.3 \cdot 10^8$ | **100** |
| $^{127}$I | 1.86 | (n,γ) | 1.317 | $6.7 \cdot 10^7$ | **14.1** |
| $^{127}$I | 1.86 | (γ,β) | 0.831 | $6.7 \cdot 10^7$ | **8.9** |
| | | | | Total | **123** |
| **HIO$_3$** | | | | | |
| $^1$H | 1.96 | (n,γ) | 2.23 | $0.74 \cdot 10^6$ | **4** |
| $^1$H | 1.96 | (n,n') | E/2 | $10^6$ | **35** |
| $^{127}$I | 1.96 | (n,γ) | (n,γ) | $1.07 \cdot 10^6$ | **16.9** |
| $^{127}$I | 1.96 | (γ,β) | 0.831 | $1.07 \cdot 10^6$ | **10.7** |
| | | | | Total | **66.6** |
| **Sb** | 4.14 | (n,γ) | 2.70 | $2.20 \cdot 10^6$ | **68** |
| $^{121}$Sb | 2.37 | (γ,β) | 1.00 | $1.58 \cdot 10^6$ | **18** |
| $^{123}$Sb | 1.77 | (γ,β) | 2.18 | $6.23 \cdot 10^5$ | **15.5** |
| | | | | Total | **101.5** |

## 3. The optimization of target dimensions

At investigation of TRIV effect in neutron-nuclei interaction one measures the number of neutrons passing through oriented nuclear target for two opposite direction of neutron spin and the value of the effect can be written

$$p_T = k l \quad (33)$$

here factor k includes nuclear characteristics, l is the target length in cm . The statistical error Δ of measurements is following

$$\Delta = 1/ ( I_o \, S \, T \, e^{-l/\lambda} )^{1/2} \quad (34)$$

here $I_o$ is the neutron density (n/cm$^2$.s), S is target area (cm$^2$), T is measurement time (d) and $\lambda$ is a neutron free path (cm).

The necessary condition for observation of the effect is $p_T / \Delta \geq 3$.

$$p_T / \Delta) = k ( I_o S T)^{1/2} l \cdot e^{-l/2\lambda} \geq 3 , \qquad (35)$$

The large number of p-resonances in the Sb$^{121}$, Sb$^{123}$ and I$^{127}$ nuclei makes it possible to perform the statistical estimation of the value of the $p_T$ effects and deformation effects in the p-resonances of the above nuclei. This estimation has been carried out in [12]. It turned out that the $p_T$ values in the p-resonances are two order of magnitude as great as the statistical error $\Delta$ obtained by TRIPLE collaboration at the measurement of P-odd effects in p-resonances at the spallation neutron source LANCE (Los Alamos, USA) [7] (i.e. $p_T /\Delta \geq 100$ at degree of alignment $p_2=1$). This comparison is appropriate since the value of the statistical errors $\Delta$ in both cases determined mainly by the width of p-resonances. Since one uses polarized neutron beam and non oriented nuclear targets at investigation of P-odd effects, TRIPLE collaboration used targets of large dimensions in their experiments : cross-section S of 85 cm$^2$ and length l of 18 cm in the case of Sb target and S=100 cm$^2$ and l=30 cm in the case of I target. The measurement time usually was 5 days. One can see from formulas (30, 31) that $p_T$ increases at increasing of target length but at the same time the statistical error also increases. So one must chose the experimental condition providing maximum value of $p_T /\Delta$ ratio. In the case of thermal neutron beam this condition is satisfied at l = 2$\lambda$. However, when one uses the resonance neutron beam and aligned targets with large number of p-resonances there is no single optimal target thickness and for each resonance the optimal length will be its own. We calculated the dependence of the value of $p_T /\Delta$ ratio on the target length l for target of I, Sb, HIO$_3$ and LiIO$_3$ single crystals. As an example the output computation for $^{121}$Sb is shown in fig. 2.

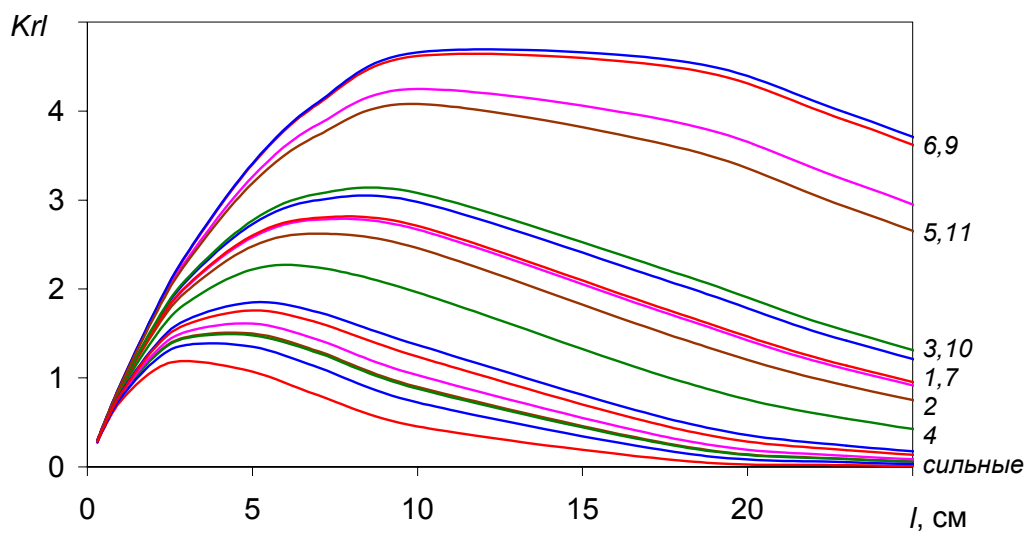

Fig. 2 Dependence of ratio $p_T /\Delta$ on target thickness for $^{121}$Sb. The numbers of the p-resonances [12] are indicated on the right. The strong resonances are: 8,12, 13, 14, 15, 16, 17.

One can see that the dependence of $p_T /\Delta$ ratio on the target length l varies significantly from one resonance to another. An optimal length for strong resonances is 3 ÷ 5 cm an for weak ones 10 ÷ 12 cm. This dependence is a similar nature for other targets. In the case of $^{121}$Sb, the value of

$p_T /\Delta$ ratio is about 1.5 at l = 5 cm and for 7 strong resonances it is five times greater than at l = 18 cm. In the case of I, the value of $p_T /\Delta$ ratio is about 3 at l = 5cm and for 9 strong resonances it is five times greater than at l = 30 cm. In the case of $HIO_3$ and $LiIO_3$ targets, the optimal length of target is also about 5 cm and for 9 strong resonances $p_T /\Delta$ ~1 at this length. By this means, one can take 5 cm as an optimal target length. For the choice of the optimal target volume one must use the data of table 4 and fig. 3 where it is shown the dependence the refrigerator cooling power Q ( at throughput of 800 μmol/s ) and nuclear alignment $p_2$ for I nuclei (in single crystals of $HIO_3$ and $LiIO_3$) and for Sb nuclei (in Sb single crystal) on temperature T.

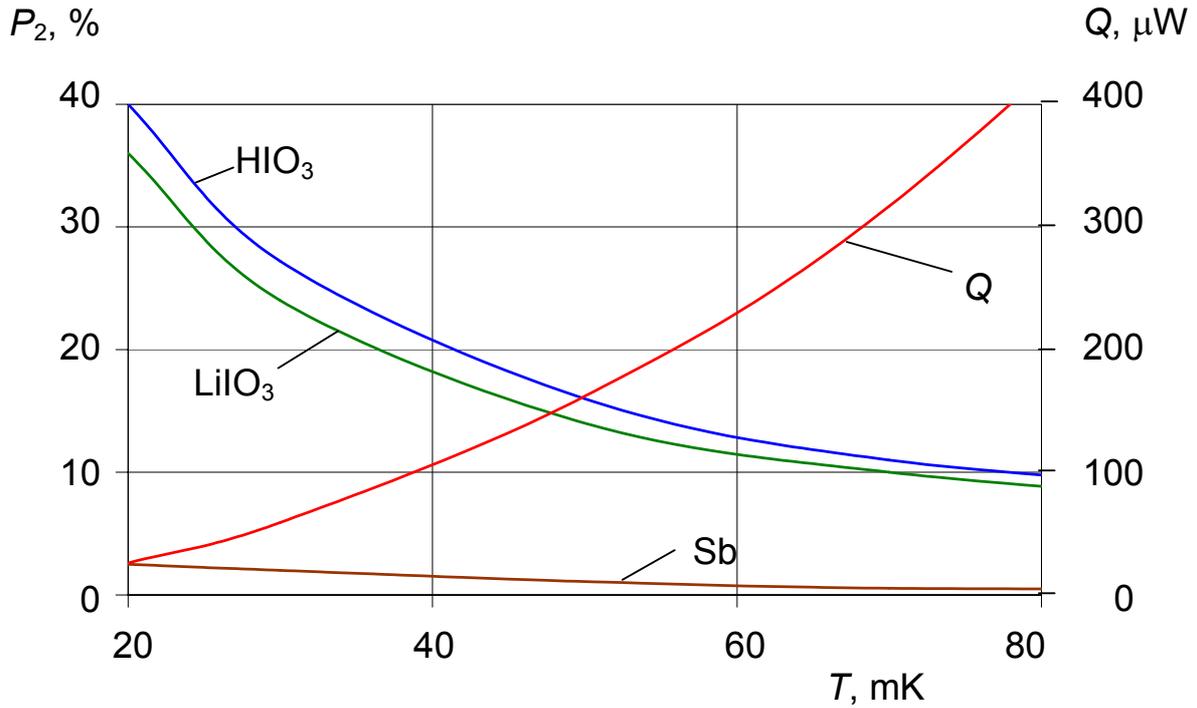

Fig. 3 Dependence of the nuclear alignment $p_2$ for I nuclei (in single crystals of $HIO_3$ and $LiIO_3$) and for Sb nuclei (in Sb single crystal) and refrigerator cooling power Q on temperature T.

It is seen from table 4 that in the case of Sb target the thermal heating is 100 μW for target volume V = 5· 5· 5 $cm^3$ = 125 $cm^3$. By reference to fig. 3 one can determine that dilution refrigerator at this value of thermal heating can cool down the target to 40 mK. The degree of the $^{121}Sb$ alignment will be $p_2$ = 1.3 % at such temperature. During the month of measurement one can achieve the limit on the value of parity conserving time violating interaction matrix element $v_T$ < 30 meV under these conditions. If one chooses the target of 50 $cm^3$ volume then the thermal heating will be smaller by a factor 2.5 , the target temperature will be about 26 mK. The degree of the $^{121}Sb$ alignment will be $p_2$ = 2 % at this temperature and the limit on $v_T$ can be decreased to 12 meV during the same time of measurement. It is evident from an analogous consideration of the case of $LiIO_3$ target that during the month of measurement one can achieve the limit on $v_T$ < 1.6 meV with the target of 125 $cm^3$ volume. If one chooses the target of 50 $cm^3$ volume then the limit on $v_T$ can be decreased to 1 meV during the same time of measurement. It should be pointed out that the present limit on $v_T$ is 1 eV [12 ].

## 4. Conclusion

Presently the way is cleared for the investigation of TRIV derived from parity conserving time violating interaction at the new neutron spallation source JSNS (Japan). The more perspective for these investigations are $Sb^{121}$, $Sb^{123}$ and $I^{127}$ nuclei having a large number of p-resonances. The abovementioned nuclei can be aligned by brute force method – by means of cooling of $LiIO_3$, $HIO_3$ and Sb single crystals to millikelvin temperatures. Now there are no technical problems for construction of aligned targets because of significant progress in crystal growth and low temperature technique. The use of proposed targets at the new neutron spallation source JSNS will make possible to discover TRIV or decrease the present limit on the intensity of parity conserving time violating interaction by two-three order of magnitude.

This work is supported by RFBR grant № 03-02-16050.